\documentclass[useAMS,usenatbib]{mnras}

\usepackage{graphicx,amsmath,amssymb}
\usepackage{hyperref}
\usepackage[T1]{fontenc}
\usepackage{aecompl}
\usepackage{blindtext}
\usepackage{natbib}
\usepackage{usebib}

\title[Working towards an optimal sampling]{Working towards an optimal sampling of the 21~cm signal parameter space}
\author[Evan Eames et al.]{Evan Eames$^{1}$\thanks{E-mail:
evan.eames@obspm.fr}, Aristide Doussot$^{1}$, Beno\^it Semelin$^{1}$.\\
$^{1}$ Sorbonne Universit\'e, LERMA, Observatoire de Paris, PSL research university, CNRS, F-75014, Paris, France}
\begin{document}

\date{Accepted XXX. Received XXX; in original form XXX}

\pagerange{\pageref{firstpage}--\pageref{lastpage}} 

\maketitle

\label{firstpage}

\begin{abstract}
With a statistical detection of the  21~cm signal fluctuations from the Epoch of Reionization (EoR) expected in the next few years, there is an interest in developing robust and precise techniques to constrain the underlying astrophysical parameters. Bayesian inference with Markov Chain Monte Carlo, or different types of supervised learning for backward modelling (from signal to parameters) are examples of such techniques. They usually require many instances of forward modelling (from parameters to signal) in sampling the parameters space, either when performing the steps of the Markov Chain or when building a training sample for supervised learning. As forward modelling can be costly (if performed with numerical simulations for example), we should attempt to perform an optimal sampling according to some principle. With this goal in mind, we present an approach based on defining a metric on the space of observables, induced by the manner through which the modelling creates a mapping from the parameter space onto the space of observables. This metric bears a close connection to Jeffreys' prior from information theory. It is used to generate a homogeneous and isotropic sampling of the signal space with two different methods. We show that when the resulting optimized samplings, created with 21cmFAST, are used to train a neural network we obtain a modest reduction of the error on parameter reconstruction of $\sim $10\%  (compared to a na\"ive sampling of the same size). Excluding the borders of the parameter space region, the improvement is more substantial, on the order of 30-40\%. 

\end{abstract}


\section{Introduction}
The cosmological $21$~cm signal, emitted in the neutral intergalactic medium (IGM) during the Epoch of Reionization (EoR), is one of the most promising observational probes of the early universe. Information is encoded in its angular and frequency fluctuations about the nature, population, distribution, and evolution of the sources of radiation at different wavelengths, but also about the cosmology of the early Universe \citep[see][for a review]{Furlanetto06}. Although a wealth of information is encoded in the signal, it remains very difficult to detect, mainly due to the high level of foreground contamination that sets stringent requirements on the necessary calibration of the instrument that aim to extract the signal \citep[see][and references therein]{Mellema13,Koopmans15}. Several instruments have been attempting to measure the power spectrum of the signal. The GMRT, MWA and LOFAR have all published upper limits \citep{Paciga13,Beardsley16,Patil17}. The PAPER team previously held the strongest uppers limits, but those have since been retracted \citep{Ali18} and new estimates are in preparation. Single dipole experiments have also been hunting for the global signal. \cite{Bowman18} have reported a possible first detection of the signal with EDGES. However, the global signal is difficult to indisputably distinguish from foregrounds and instrumental effects, and even though the EDGES detection has revealed an unexpectedly sharp feature in frequency, this must now be confirmed with interferometric observations. Over the coming decade, HERA and the SKA are expected to be able to measure the power spectrum of the signal with high accuracy and, for the latter, build a full tomography.

The local intensity of the signal is a function of the neutral hydrogen number density, the peculiar velocity of the gas, and, through the spin temperature, also of the kinetic temperature of the gas and the local Lyman-$\alpha$ flux. These quantities result from non-local processes (gravitation, radiative transfer), and are non-trivially correlated. That is to say, extracting the full information encoded in the signal remains a difficult task. The most realistic approach is to model the signal based on a small set of parameters, using either theoretical, semi-numerical, or full-numerical methods, and then to infer constraints on these model parameters based on real observations. One difficulty is that there is, at present, no unique and well-established set of quantities with which to parametrize the underlying (unresolved) astrophysical processes that shape the signal. A number of possible parameters have been proposed \citep[see e.g.][]{Greig15,Cohen17,Greig17, Semelin17}.  

Regardless of the chosen set of parameters, the next step is to develop a method with which to determine the parameter values that, when used for simulating the signal, result in a model that most closely matches observations according to some metric, as well as (if possible) confidence intervals that quantify the uncertainty due to the thermal noise present in the observed signal.

The Fisher information matrix, which can be used as a tool for finding the maximum likelihood (e.g. Markov Chain Monte Carlo, MCMC henceforth) has long been made use of in astrophysics. It has been applied to the 21~cm signal by \cite{Pober14} to derive confidence contours, while not explicitly used in the search for the maximum likelihood of parameter values. Another widely used method is Bayesian inference, again making use of MCMC. It typically requires many instances of forward modelling, and can also be used to provide confidence contours. In the context of the 21~cm signal, Bayesian inference has been applied using semi-numerical codes as the model \citep[namely 21cmFast, see][]{Greig15,Greig17,Greig18}, or, to speed up the process, using emulators based on Gaussian Processes \citep{Kern17}, or neural networks \citep{Schmit18}. Finally, parameter predictions (without constraints in terms of confidence levels) can be obtained through machine learning. For example, \cite{Shimabukuro17} and \cite{Gillet18} train neural networks to perform backward modelling, and thus recover the model parameters from a signal not used in the training. It is worth noting that the size of the training sample required by neural networks and created with forward modelling tends to be much smaller than the number of Markov Chain steps (also instances of forward modelling) in Bayesian MCMC inference. Therefore, there is hope that supervised learning in general, and in particular neural network training, could be performed using full-numerical simulations at some point.

Whether choosing a prior with which to perform Bayesian inference, or when building a training sample for supervised learning techniques, the choice of a distribution in the parameter space is crucial. A flat distribution (or `flat prior') on the parameters (or their logarithm) for Bayesian inference is a natural heuristic first approach, as is a grid-like uniform sampling of the parameter space for supervised learning. However, more informed choices are possible such as the Jeffreys' prior for Bayesian inference \citep{Jeffreys46}. This prior is such that it generates a flat distribution of the {\sl noisy} observable quantity, and thus is much more agnostic regarding the model than a flat prior on the parameter distribution. The difficulty is that the full knowledge of the Fisher information matrix (that is, knowledge about both sensitivity to the parameters and noise effects at every point in the parameter space) is required to compute it. One can expect that a homogeneous and isotropic sampling in the space of the observable quantity, corresponding to the flat distribution created with Jeffreys' prior, would also optimize the process of supervised learning for backward modelling. In this work, we make a first step in this direction. We quantify a metric equivalent to the Fisher information metric in the case where thermal noise is not included (and which can be directly equated to the Fisher information matrix in some cases). Such a metric quantifies the sensitivity to the parameters, but does not take into account the variance of the stochastic process of thermal noise. Our goal is to show how a training sample built using this metric allows for more efficient supervised learning in the case where a given neural network is trained from signals not including thermal noise. Adding thermal noise and using the Fisher information metric will be the next step, in an upcoming article.

In Section 2 we briefly set up some definitions that will be useful for properly explaining the procedure. Section 3 outlines the parameter space upon which this process is carried out, the simulation preliminaries required to proceed, and the anisotropies/inhomogeneities we look to address. Section 4 explains a first, global, methods to correct it. Section 5 presents a second algorithm that takes into account the local values of the metric. In Section 6 we apply these findings to training a neural network in parameter reconstruction, to see if the optimal parameter space is truly more efficient. Finally, Section 7 summarizes these results and looks towards future work.

\section{Terminology}
Here we allow ourselves the liberty of defining some useful terms. Some are rather obvious while others are sometimes ambiguous (like `model') or specific to our work.

\noindent \underline{The parameter space} is an $n$-dimensional space where each dimension corresponds to the value of a different parameter.

\noindent \underline{A sampling} is a finite choice of $k$ points within a given $n$-dimensional parameter space.

\noindent \underline{An observable} is a quantity that can be directly derived from an observation. The power spectra, pixel distribution functions, lightcones, the global signal, etc. are all choices of observables for $21$~cm observations.

\noindent \underline{The space of observables} is the space spanned by all possible values of the chosen observable. For example, a power spectrum estimated in 10 $k$-bins at a single redshift inhabits a 10-dimensional space of observables.

\noindent \underline{A model} is a framework (theoretical or numerical) that is used to compute an observable for any given point in the parameter space.

\noindent \underline{The hypersurface of predictions} by the model is a manifold embedded in the space of observables. The model acts as a map between the parameter space and this hypersurface, both of which have the same dimension. The geometry of this hypersurface can be quite different from that of the parameter space. For example, the two closest points in a sampling may not necessarily transform to give the two closest observables. This of course depends of the definitions of distances in both spaces.

\noindent \underline{An optimal sampling} is, as per our definition, a sampling of the parameter space that maps onto a homogeneous and isotropic sampling of the hypersurface of predictions. Should the points in parameter space be organized on a grid, we would ideally like for any two neighbouring points (along any axis) to map to equally different observables (in the sense that they are at equal distance for a given distance definition in the observable space). If the sampling is not performed on a grid, the definition of neighbouring points is less straightforward and we will return to this case in section \ref{Adaptive Grid-free Method of Resampling}. The specific configuration of an optimal sampling of the parameter space will depend on both the chosen observable and the definition of distance in the space of observables. If we define the distance to be the L$_2$ norm for power spectra weighted by the inverse variance, the definition of optimal directly derives from the Fisher information metric (see section \ref{Metric Definition}).

\section{Methods} \label{Methods}
The procedure for developing and testing an algorithm that creates an optimal sampling requires many repetitions of observable prediction (for a choice of model). At present it is computationally infeasible to run a full-numerical simulation this many times. For this reason, we have used a semi-numerical model instead: 21cmFAST\footnote{http://github.com/andreimesinger/21cmFAST} \citep{Mesinger07, Mesinger11} provides the required speed and efficiency.

\subsection{Parameter Space}\label{Explored Parameters}
Our chosen parameter space consists of three parameters that have been explored previously by other authors \citep[e.g.][]{Greig15,Greig17,Greig18,Schmit18}. These parameters are:

\begin{itemize}
\item $\zeta$, the ionizing efficiency of high-z galaxies:
\begin{equation}
\zeta = 30 \left(\frac{f_{esc}}{0.3}\right)\left(\frac{f_\star}{0.05}\right)\left(\frac{N_\gamma}{4000}\right)\left(\frac{2}{1+n_{rec}}\right)
\end{equation}
where $f_{esc}$ is the ionizing photon escape fraction, $f_\star$ is the fraction of galactic gas in stars, $N_\gamma$ is the number of ionizing photons produced per baryon in stars, and $n_{rec}$ is the typical number of times a hydrogen atom recombines during the EoR.
\item $R_{\textrm{mfp}}$, the mean free path of ionizing photons within ionized regions, set by the existence of unresolved Damped Lyman-$\alpha
$ systems.
\item $T_{\textrm{vir}}$, the minimum virial temperature for halos to be allowed to form stars. It can be thought of as an ionization switch, controlling when growing halos begin to ionize their surroundings. Conversely to $\zeta$, setting it to high values will result in a neutral universe up to very late redshifts.
\end{itemize}
Full definitions are given in \citet{Greig15}. For our purposes, we have assumed the following ranges for these three parameters:
\begin{itemize}
\item[-] $\zeta \in [20,200]$
\item[-] $R_{\textrm{mfp}} \in [5 \,\mathrm{cMpc},35 \,\mathrm{cMpc}]$
\item[-] $T_{\textrm{vir}} \in [8\times10^3\,\mathrm{K}, 10^5\,\mathrm{K}]$
\end{itemize}
A preliminary set of `test models' showed that those created outside of these ranges could exhibit unwanted behaviour, such as a complete lack of reionization by $z = 6$.

\subsection{Simulation Configuration} \label{Simulation Configuration}
We created a wrapper to run many instances of 21cmFAST in parallel. The parameter file was altered as required for each clone, in order to vary the $\zeta$, $R_{\textrm{mfp}}$, and $T_{\textrm{vir}}$ values as needed. 21cmFAST version 1.2 was used, with a box size of 300 cMpc, a high resolution (for sampling initial conditions) of 768 cells per axis, a low resolution (for evolving the box) of 256 cells per axis, and outputs were generated starting at $z_{max} = 15$, 	at intervals of $\Delta z = 1$. We also set $T_S \gg T_{\textrm{CMB}}$ which is valid once the heating of the neutral IGM by X-ray sources has had enough time to operate. In most scenerios this is valid for most of reionization, although the assumption is expected to be incorrect during the Cosmic Dawn \citep[e.g.][]{Baek10,Fialkov14a}. For our purposes, assuming $T_S \gg T_{\textrm{CMB}}$ results in a speed-up of $\sim$ a few, and we remind the reader that our goal is to study sampling optimization, not the effects of different heating scenarios.

The code was run on the OCCIGEN supercomputer, maintained at CINES\footnote{\href{https://www.cines.fr/calcul/materiels/occigen/}{https://www.cines.fr/calcul/materiels/occigen/}}. For a sampling of 1,000 points, the runtime is approximately a few hours on OCCIGEN, using one core per instance of the code. We chose the power spectra $P(k,z)$ as our observable. We define the distance between two observables to be the same as in \citet{Semelin17}, based on the L$_2$ norm:

\begin{equation} \label{equation:distance}
\textrm{D}_{i,j} = \sqrt{\int(\textrm{P}_i(k,z) - \textrm{P}_j(k,z))^2dkdz}
\end{equation}

Where P$_i$ and P$_j$ are the power spectra of two points in our sampling, functions of wavenumber $k$ (in cMpc$^{-1}$) and redshift $z$. Again, see section \ref{Metric Definition} for a discussion on a different definition that would match the Fisher information metric, and an explanation of why we do not adopt it in this work.

\section{Properties of a na\"ive Logarithmic Sampling} \label{Initial Logarithmic Sampling}

We first create a na\"ive `fiducial' sampling and use it to quantify the geometrical properties of the hypersurface of predictions. We will then use this information to produce samplings closer to optimal. We choose to start with a logarithmic sampling. As a first test, we evaluated the `density' of the mapping of our sampling onto the hypersurface of predictions, using the distance defined in equation \ref{equation:distance}. An initial 8$\times$8$\times$8 sampling is used for this preliminary study.

\subsection{Inhomogeneity and Anisotropy in the Parameter Space} \label{Inhomogeneity and Anisotropy in the Parameter Space}
\begin{figure}
\includegraphics[width=0.44\textwidth]{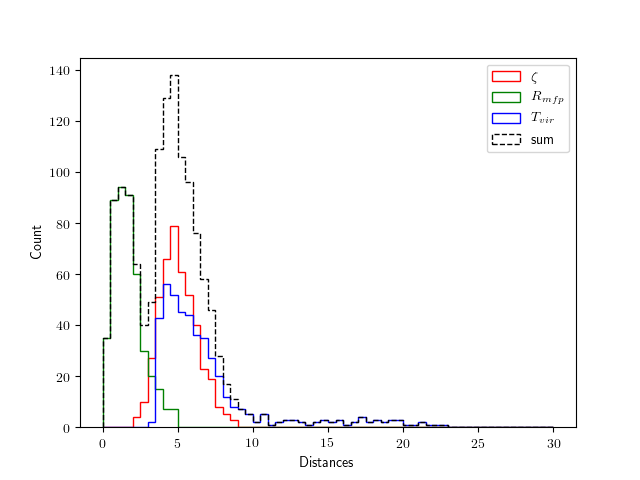}
\caption{Distances between neighbouring models along the three parameter axes.}
\label{fig:Distance_histo1}
\end{figure}

We can easily explore the statistics of distances between the observables predicted by the models, starting first with the distances between neighbouring points of this fiducial grid sampling of the parameter space. This entails the distance between each point and its six grid neighbours. Examining the distribution of these distances (studied separately along each parameter axis in figure \ref{fig:Distance_histo1} for the 8$\times$8$\times$8 sampling), we see that our simplistic logarithmic sampling maps onto an inhomogeneous and anisotropic distribution on the hypersurface of predictions, when defining distance as the L$_2$ norm on the power spectra. 

We know this because an isotropic distribution would have identical histograms for each axis, which is not the case here. As for a homogeneous distribution, this would correspond to Dirac delta functions (not peaking necessarily at the same values if isotropy is not satisfied). We do not see this here: the $R_{\textrm{mfp}}$ axis distances are smaller than along the other two axes, all three histograms have some width, and the $T_{\textrm{vir}}$ histogram has some outlier values scattered around at higher distances. An optimal sampling would therefore be three Dirac delta functions centred at the same value.

To progress towards this goal, we need to quantify the geometry of the hypersurface of prediction in more detail. We can achieve this by evaluating a metric associated with our distance definition. The initial 8$\times$8$\times$8 sampling proved to be too sparse for finite difference metric estimation (see section \ref{Metric Definition}). The sparsity was especially damaging in two of the three dimensions due to the  strong anisotropy. Hence, for the estimation of the metric, we adapt the fiducial sampling to consists of 2,400 points, corresponding to $20\times6\times20$ ($R_{\textrm{mfp}}$ is sampled for 6 values), spaced logarithmically between the ranges presented in section \ref{Explored Parameters}. We could simply have used a much finer initial sampling but this would have been more costly.

\subsection{A Metric on the Hypersurface of Predictions} \label{Metric Definition}

\subsubsection{Metric definition}
A metric allows us to convert the coordinate separation of two points into a distance between those two points. Thus we fist need a coordinate system for the hypersurface of predictions. Assuming that the model operates a bijection between the parameter space and the hypersurface of predictions (which will be true in non-pathological cases, at least within a finite region), we can simply use the parameter values as a coordinate system for the hypersurface of predictions. For our purpose we use the log$_{10}$ of the parameter values, and normalize each axis so that each of the three logarithmic steps of our initial sampling is assigned a norm of $1$. Thus, for our choice of coordinate system, the points of our initial sampling are located at all points with integer-only coordinates in a parallelepiped rectangle of size $20 \times 6 \times 20$. We will denote the three coordinates associated to the parameters  $\theta_1$, $\theta_2$, $\theta_3$.

Let us now denote the metric of the 3D hypersurface of predictions as:

\begin{equation}
\mathbf{g} = 
\begin{bmatrix}
    g_{\theta_1\theta_1} & g_{\theta_1\theta_2} & g_{\theta_1\theta_3} \\
    g_{\theta_1\theta_2} & g_{\theta_2\theta_2} & g_{\theta_2\theta_3} \\
    g_{\theta_1\theta_3} & g_{\theta_2\theta_3} & g_{\theta_3\theta_3} 
\end{bmatrix}
\end{equation}
such that two points in the hypersurface of prediction separated by an infinitesimal vector $\begin{bmatrix}d\theta_1&d\theta_2&d\theta_3\end{bmatrix}$ (using the above coordinate system), will be separated by a distance dl given as:
\begin{equation}\label{eqn:distance}
\textrm{dl}^2 =
 \begin{bmatrix}d\theta_1 & d\theta_2 & d\theta_3 \end{bmatrix} \cdot \mathbf{g}\cdot 
 \begin{bmatrix}d\theta_1 \\ d\theta_2 \\ d\theta_3 \end{bmatrix}
\end{equation}

\subsubsection{Computing the metric}
We now need a numerical scheme with which to compute the metric at each point of our logarithmic sampling. We use a simple finite difference scheme. In our coordinate system, the vectors between neighbouring grid points can be written $\begin{bmatrix}\Delta_{\theta_1}&\Delta_{\theta_2}&\Delta_{\theta_3}\end{bmatrix}$ where each of the $\Delta$ terms can be equal to $-1$, $0$, or $1$. We are thus considering the 26 neighbours defined by a cube centred on the point at which we are calculating the metric. Let $D_{\Delta_{\theta_1},\Delta_{\theta_2},\Delta_{\theta_3}}$ be the corresponding distance according to the metric, we have the following relations for the vectors in the $(\theta_1,\theta_2)$ plane:

\begin{eqnarray}
D^2_{1,0,0} &=&g_{\theta_1\theta_1} \\
D^2_{-1,0,0} &=&g_{\theta_1\theta_1} \\
D^2_{1,1,0} &=&g_{\theta_1\theta_1}+2g_{\theta_1\theta_2}+g_{\theta_2\theta_2} \\
D^2_{-1,-1,0} &=&g_{\theta_1\theta_1}+2g_{\theta_1\theta_2}+g_{\theta_2\theta_2} \\
D^2_{1,-1,0} &=&g_{\theta_1\theta_1}-2g_{\theta_1\theta_2}+g_{\theta_2\theta_2} \\
D^2_{-1,1,0} &=&g_{\theta_1\theta_1}-2g_{\theta_1\theta_2}+g_{\theta_2\theta_2} 
\end{eqnarray}
Twelve more equations can be written when we include the $(\theta_1,\theta_3)$ and $(\theta_2,\theta_3)$ planes. This constitutes an overdetermined set of equations (which could be further expanded to include the corners of the cube) for which we could easily find the least mean squared approximate solution. The symmetries of the equations suggest a simpler scheme. We use the following as an approximate solution:

\begin{eqnarray}
g_{\theta_1\theta_1}&=&{D^2_{1,0,0} + D^2_{-1,0,0} \over 2} \\
 g_{\theta_1\theta_2}&=&{ D^2_{1,1,0} +D^2_{-1,-1,0} -D^2_{1,-1,0} -D^2_{-1,1,0} \over 8} \
\end{eqnarray}
along with equivalent expressions for the other coefficients. This scheme can be thought of as similar to computing a derivative with a centred scheme. Because of this choice, we do not compute the metric at points located on the faces of our parallelepiped domain. The accuracy of this finite diference estimation obviously depends on our fiducial sampling being dense enough so as to prevent the metric from experiencing high variation between any two neighbouring points.

\subsubsection{Link with the Fisher information metric}

If we include the contributions of sample variance and thermal noise, the model for computing a power spectrum for a given set of parameter values can be viewed as a stochastic process. Then the distribution $f$ of possible power spectra binned at wavenumbers $k_i$ and redshifts $z_j$ is:

\begin{equation}
f\left(\,P_N, \vec{\theta}\right)\sim \prod_{k_i,z_j} \exp \left[ { (P_N(k_i,z_j) - P(k_i,z_j, \vec{\theta}) )^2\over 2(\sigma(k_i,z_j))^2 } \right]
\end{equation}
where $P_N$ is the noisy power spectrum, $\vec{\theta} = \begin{bmatrix}\theta_1&\theta_2&\theta_3& ...& \theta_n\end{bmatrix}$ are the parameters of the model, and $\sigma^2$ is the variance of the combined sources of noise. We have assumed Gaussian noise, which is valid if the bins are sufficiently large (a consequence of the central limit theorem). We have also assumed that the thermal noise is uncorrelated at different wavenumbers (which requires non overlapping bins, but ignores possible instrumental effects). Then the distribution for the full power spectrum is simply the product of the distributions for each bin. We can now inject this formula into the definition of the Fisher information matrix coefficients:

\begin{equation}
I_{\theta_1,\theta_2} = E \left[ {\partial \log f \over \partial \theta_1}  {\partial \log f \over \partial \theta_2} \right]
\end{equation}
\begin{equation}
= \sum_{k_i,z_j} {1 \over (\sigma(k_i,z_j))^2} {\partial P(k_i,z_j,\vec{\theta}) \over \partial \theta_1} {\partial P(k_i,z_i,\vec{\theta}) \over \partial \theta_2} 
\end{equation}
where $E$ designates the expectation of the distribution. These are the coefficients of the matrix representation of the Fisher information metric. We can establish a similar expression for our metric $\mathbf{g}$. Using a version of eq. \ref{equation:distance} discretized on the wavenumber and redshift bins, applying it to a pair of power spectra differing by infinitesimal variations of the parameters, using a first order Taylor expansion in the parameters, and comparing with eq. \ref{eqn:distance}, one gets:

\begin{equation}
g_{\theta_1 \theta_2}=\sum_{k_i,z_j} {\partial P(k_i,z_j,\vec{\theta}) \over \partial \theta_1} {\partial P(k_i,z_j,\vec{\theta}) \over \partial \theta_2} 
\end{equation}

Of course we arrive at a similar expression for any pair of parameters. We can also see that the two expressions differ only by the inverse variance of the noise term, appearing only in the Fisher information matrix. In some sense the Fisher metric defines a more meaningful distance which accounts for the rms noise. We can also see that this meaningfully modifies the metric only if the level of noise is different in different $k$-bins (which is indeed the case). Also, one can note that it is a simple matter to match the definition of the Fisher metric in our procedure for computing the metric; we need only to plug in the power spectra weighted by the inverse of the rms noise level (weighted for each $k$ and $z$ bin).

The reason why we do not use the Fisher metric is that, first and foremost, we are interested in exploring an optimal sampling --- ideally with full-numerical simulations (should this become feasible in future) --- to use as training data for neural networks. At this stage, as shown in Section \ref{sec:NN}, it has been shown that the improvement gained through an optimal learning sample is only measurable for a network trained with noise-free `idealized' power spectra as training input. In the case of networks trained with noisy power-spectra, we feel that the error in the recovered parameter values may still be dominated by other factors in training the network, and this would ultimately negate any improvement brought on by the optimal sampling. We intend to explore the Fisher metric optimal sampling in a future work, so as to both apply a Jeyfreys prior to Bayesian MCMC inference, or to use with improved networks trained on noisy data.

\subsection{Eigenvectors and Eigenvalues} \label{Eigenvectors and Eigenvalues}

Strictly speaking, because a metric is a tensor, there is technically no true definition of eigenvectors or eigenvalues. This being said, we are still free to consider the matrix corresponding to said metric, and to diagonalize it. The eigenvector with the largest (resp. smallest) eigenvalue will be the direction in the parameter space where a coordinate displacement of norm $1$ will produce the largest (resp. smallest) distance between the corresponding observables. As such the eigenvectors and eigenvalues contain information on how the geometry of the hypersurface of observables relates to the chosen coordinate system. This information can help us to produce an improved sampling. It should also be noted that the eigenvectors are each a combination of the three parameter vectors ($\zeta$, $R_{\textrm{mfp}}$, and $T_{\textrm{vir}}$), and there is no direct correspondence between the distances in figure \ref{fig:Distance_histo1} and the eigenvalues.

\begin{figure*}
\minipage{0.48\textwidth}
  \includegraphics[width=\linewidth]{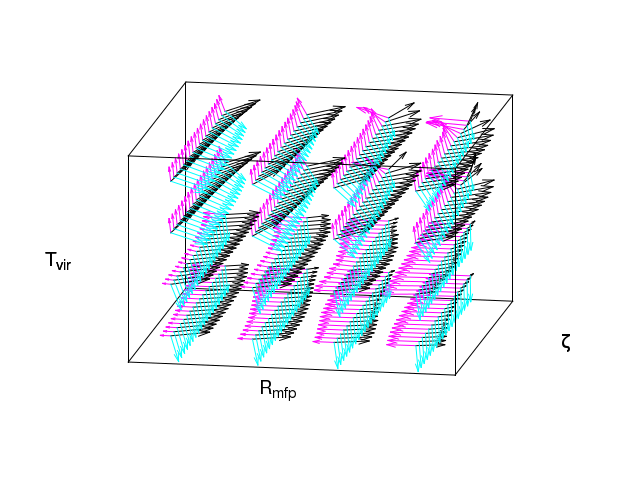}
\caption{The eigenvectors at each point within a region of the parameter space, showing how they appear to `rotate' from one point to the next along the axes.} \label{fig:EigenvectorField}
\endminipage\hfill
\minipage{0.48\textwidth}%
\centering
  \includegraphics[width=0.82\linewidth]{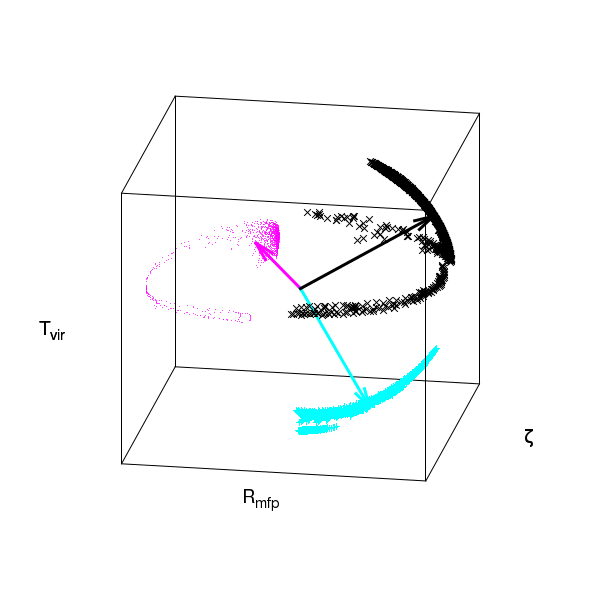}
\caption{Here we have superimposed the heads of all eigenvectors. The clusters represent the heads of the individual eigenvectors (such as those in figure \ref{fig:EigenvectorField}), and the thick arrows represent the averaged eigenvectors.} \label{fig:Eigenvectors}
\endminipage
\end{figure*}

We can explore the distributions of the eigenvectors, which are summarized in figure \ref{fig:Eigenvectors}. We see that for our sampling, the corresponding eigenvectors tend to fall in clustered regions (pointed to by the averaged eigenvectors). However, the eigenvectors $v_2$ and $v_3$ have `tails' wherein the values spread out. These correspond to regions of the parameter space where the geometry of the hypersurface of predictions changes quickly.

\subsubsection{Eigenvector matching}
In diagonalizing the matrix corresponding to the metric at each point, a problem arises. As the distance between two neighbouring points is not infinitesimal (for our finite sampling), there can be situations where the hypersurface changes rapidly and it is therefore not obvious how a given eigenvector would be associated to another one at the neighbouring point if a continuous path was followed. Moreover, $\vec{v}_i$ and $-\vec{v}_i$ are equally valid eigenvectors (with eigenvalue $\lambda_i$), and the diagonalization routine\footnote{We use the DSYEVJ3.c script, which diagonalizes using the Jacobi method \citep{Kopp08}.} will sometimes switch between signs at arbitrary positions in the parameter space to satisfy accuracy requirements. To apply the methods described below (using average eigenvectors), we need to consistently group the eigenvectors into 3 matching families.

To achieve this, we compare each point to its neighbours, sort the three vectors by minimizing angle separations, and see if any of them have been inverted. For especially difficult regions, we also use the fact that eigenvalues and eignevectors should exhibit a regularity in their evolution across the parameter space. Take for example the rotation we see across the parameter space (figure \ref{fig:EigenvectorField}), which can be used to extrapolate where the eigenvectors of the next point are expected to be.

\subsection{Using Average Eigenvectors for Resampling} \label{Grid Reorientation}
Computing the 3 average unitary eigenvectors $\bar{v}_n$ across the parameter space and their corresponding average eigenvalues $\bar{\lambda}_n$, we effectively know the `average' distance between simulations (eigenvalues) when travelling through the parameter space in three orthogonal directions (eigenvectors). We can therefore use this knowledge to re-sample the parameter space such that, in the new grid, the distances between neighbouring simulations will be closer to constant on average. It is to be expected that, on account of regions where the eigenvectors experience rotations away from the average, there will be some variation in the distances between neighbouring observables in said regions. However, we should still expect a more isotropic and homogeneous parameter sampling than the logarithmic counterpart.

\begin{figure*}
\minipage{0.32\textwidth}
  \includegraphics[width=\linewidth]{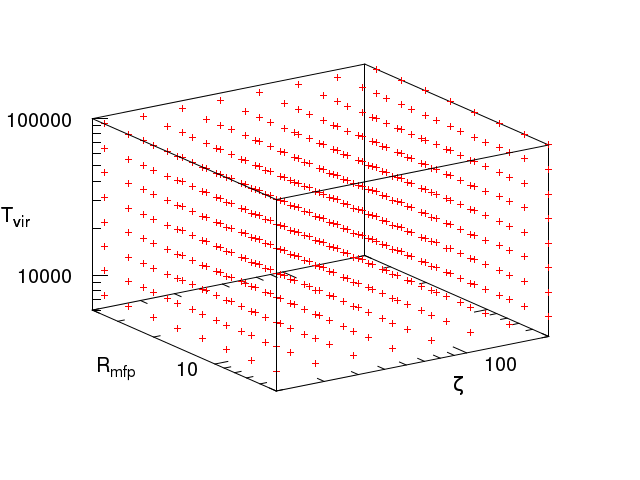}
    \centering
  \textbf{(a)} Logarithmic sampling
\endminipage\hfill
\minipage{0.32\textwidth}
  \includegraphics[width=\linewidth]{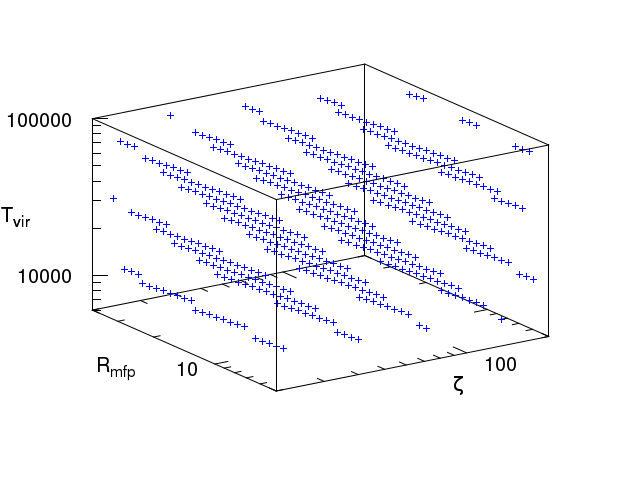}
    \centering
  \textbf{(b)} Average-eigenvector sampling
\endminipage\hfill
\minipage{0.32\textwidth}%
  \includegraphics[width=\linewidth]{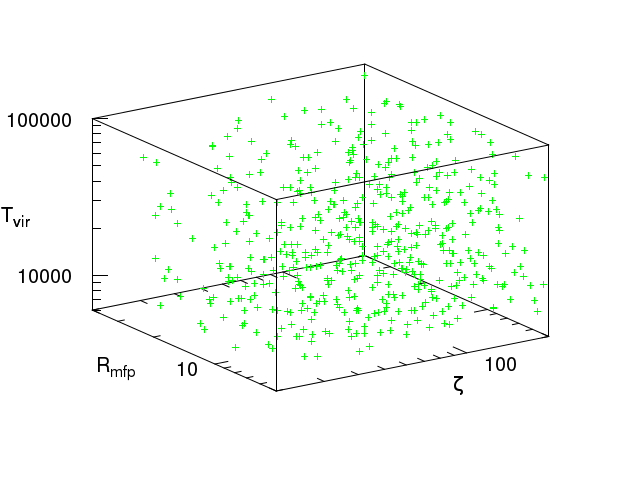}
  \centering
  \textbf{(c)} Adaptive grid-free sampling
\endminipage
  \caption{Visual representation of the three different samplings of the 3D parameter space. Each dot represents one sample that consists of triplet of values of the three parametres.}\label{fig:visualized samplings}
\end{figure*}

Starting from the central point of our parameter space, we can expand outwards using three new `step vectors', which are to be based on the average eigenvectors and their average eigenvalues. To begin, a point $i$ in our parameter space can be assigned a `volume' in the space of observables, that is to say, a region closer to this point than to any other (again based on the $L_2$ norm distance):
\begin{equation}
\label{equation:volume}
V_i = \sqrt{\textrm{det}_i} = \sqrt{\lambda_1^i \lambda_2^i \lambda_3^i}
\end{equation}
where $\lambda_n^i$ is the $n^{th}$ eigenvalue for point $i$. Assuming $N$ points in our parameter space, we can calculate the average volume to be $\bar{V} = \frac{1}{N}\sum_i V_i$, and therefore the average distance between points will be\footnote{To be clear, this is again a simplification. The true average distance depends on the geometry of the sampling (Cartesian grid, crystal lattice, etc.).}:
\begin{equation}
\bar{d} =  \sqrt[3]{\bar{V}} = \sqrt[3]{\frac 1N \sum_i\sqrt{\lambda_1^i \lambda_2^i \lambda_3^i}}
\end{equation}
We can now use the normalized average eigenvectors $\bar{v}_n = \begin{bmatrix} \bar{\theta}_{1,n} &\bar{\theta}_{2,n} &\bar{\theta}_{3,n}\end{bmatrix}$ (linear combinations of the three normalized initial parameter axis vectors $\theta_1$, $\theta_2$, $\theta_3$) and their corresponding eigenvalues $\bar{\lambda}_n$. Starting in the centre, we wish to move along each eigenvector by some distance such that the $L_2$ norm distances of the resulting observables are equal to the average distance. We know that the amount we should move along each eigenvector depends on that eigenvector's eigenvalue, so let us define a normalization constant that depends on the relevant eigenvalue: $\alpha_n(\bar{\lambda}_n)$. Now, let us take the first average eigenvector $\bar{v}_1$. If we move from the central point to a new point along the vector $\alpha\bar{v}_1$ the distance between the two corresponding observables will be (from equation \ref{eqn:distance}):
\begin{equation}
d^2 = \alpha_1\bar{v}_1 \cdot \mathbf{g} \cdot \alpha_1\bar{v}_1 = \bar{\lambda}_1 \alpha_1^2 \left(\bar{\theta}_{1,1}^2+\bar{\theta}_{2,1}^2+\bar{\theta}_{3,1}^2\right) = \bar{\lambda}_1 \alpha_1^2 
\end{equation}
To assure that $d = \bar{d}$ we can now set $\alpha_n(\bar{\lambda}_n) = \frac{\bar{d}}{\sqrt{\lambda_n}}$. The final step is to convert from the current coordinate system which normalized the logarithmic steps to one, back to the physical value of the parameters.
 We define the constants: $c_1 = \Delta \log{\zeta} ,c_2 = \Delta \log{R_{\textrm{mfp}}}, c_3 = \Delta \log{T_{\textrm{vir}}}$ (the difference between the logarithms of neighbouring parameter values). Finally, we can define step vectors $\vec{s}_n$:
\begin{equation}
\vec{s}_n = \left( \frac{\bar{d}}{\sqrt{\bar{\lambda}_n}}c_1\bar{\theta}_{1,n}, \frac{\bar{d}}{\sqrt{\bar{\lambda}_n}}c_2\bar{\theta}_{2,n},\frac{\bar{d}}{\sqrt{\bar{\lambda}_n}}c_3\bar{\theta}_{3,n}\right)
\end{equation}
This formula will give us three step vectors to be used on the logarithm of the parameters. Starting from the central point we can move outwards along said vectors, until we find ourselves outside the initial bounds defined in section \ref{Explored Parameters}, to re-sample our space. Should we require a specific number of points in our new sampling, we can adjust the $\bar{d}$ value until we have the desired number (a smaller $\bar{d}$ will result in more points, and vice versa). A re-aligned sampling with 512 points is shown in figure \ref{fig:visualized samplings} (b).



\begin{figure}
\centering
\includegraphics[width = 0.5\textwidth]{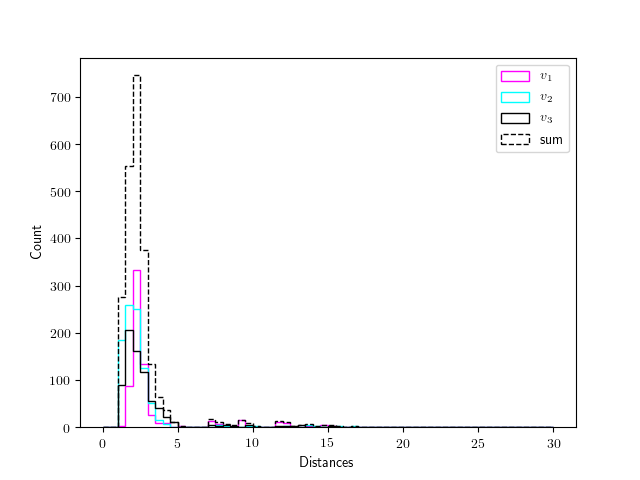}
\caption[Histograms of the distances between neighbouring observables after having resampled with the average-eigenvalue method.]{Histograms of the distances between neighbouring observables after having been resampled with the average-eigenvalue method (for the distances before sampling, see figure \ref{fig:Distance_histo1}).}
\label{fig:Resampled_Histos1}
\end{figure}

As we see in figure \ref{fig:Resampled_Histos1}, the eigenvector method for resampling the parameter space is clearly an improvement over the initial sampling. In particular, the similarity of observables along the $R_{\textrm{mfp}}$ axis (in figure \ref{fig:Distance_histo1}) has been remedied by reorienting the grid, such that no axis corresponds to a change in only $R_{\textrm{mfp}}$. As well, regions in which observables change rapidly (the $T_{\textrm{vir}}$ tail at distance $\gtrsim 10$ in figure \ref{fig:Distance_histo1}) are nearly all taken care of after resampling. Yet there is still room for improvement. Along all three (new) axes there are slight tails at higher distances in the histogram.

\section{Adaptive Grid-free Method of Resampling} \label{Adaptive Grid-free Method of Resampling}

Using an average of the metric at each point (or rather, their corresponding eigenvalues and eigenvectors) was motivated by the ease of creating a sampling on a grid. However, if we abandon the requisite of a grid altogether, we can further reduce the inhomogeneity and anisotropy by using the local information given by the metric. We now present an algorithm to attempt this.

\medskip
\noindent
{\bf Initialization}

\noindent
Again starting with the logarithmic sampling, the metric at each point is computed through the same equations defined in section \ref{Metric Definition}. The total volume\footnote{Or hypervolume.} of the parameter space is computed using the metric. Assuming maximum n-sphere packing, the typical volume for each n-sphere is calculated. We initiate a maximum interaction distance equal to twice the radius of the n-spheres ($\textrm{D}_{max}$). Along the extremities of the parameter space, we also designate a `buffer zone' of set width. A new sampling is created with a flat, random distribution in the accepted parameter region (points can also fall into a preset `buffer zone' running along the boundaries of the space).

\medskip
\noindent
{\bf Iterating}
\begin{itemize}
\item The metric is computed at each point of the sampling by interpolation using the values of metrics at the neighbouring points in the fiducial grid sampling. Interpolation is carried out with a Gaussian kernel smoothing.
\item For all pair of points $i$ and $j$:
\begin{enumerate}
\item Average the metric between these two points.
\item Use this averaged metric to approximate the distance between the observables at these two points ($\textrm{D}_{i,j}$).
\end{enumerate}

\item For each pair of neighbouring points $(i,j)$, if $\textrm{D}_{i,j} < \textrm{D}_{max}$ then we define a displacement:
\begin{equation}
\vec{d}_{i,j} = -\frac{1}{2}(\textrm{D}_{max} - \textrm{D}_{i,j})\cdot\vec{r}_{i,j}
\end{equation}
where $\vec{r}_{i,j}$ is a unitary vector pointing from $i$ to $j$.

\item For each point $i$, the displacements induced by the $n$ nearby points are averaged, and then the point is moved along this new vector $\vec{d}_i = {1 \over n}\sum_{j=1}^n \vec{d}_{i,j}$.

\item Points in the buffer zone receive an additional displacement towards the centre (representing a sort of confinement).

\item A new interaction distance $\textrm{D}_{max}$ is evaluated using the current number of points in the parameter space region. The above steps are then repeated.

\end{itemize}

\medskip
\noindent
{\bf Final Sampling}

\noindent
The positions of the points after each iteration are recorded. At the end of the loop, the best configuration can be selected from the histogram of the distances (computed using the interpolated metrics between pairs of points) from each point to the $N_{kiss}$ points, where $N_{kiss}$ is the `kissing' number for a given dimension\footnote{The kissing number is defined as follows. For a given n-sphere, how many n-spheres of the same radius can be made to touch the first one without overlaps. For 2D $N_{kiss}=6$, for 3D $N_{kiss}=12$, etc. The kissing number is not known exactly for dimension $>4$, except for 8D and 24D, however bounds have been set at $\sim20$\% up to 24D \citep{Mittelmann09}.}.

Figure \ref{fig:visualized samplings} (c) shows an example of the best sampling after a number of iterations. It's important to note that, although the sampling was initialized with 512 points, there are \emph{not} 512 points in the final sampling (but rather 506). In our current method, the adaptive grid-free algorithm generally will not preserve the number of points in the chosen parameter space region.

\begin{figure}
\centering
\includegraphics[width = 0.5\textwidth]{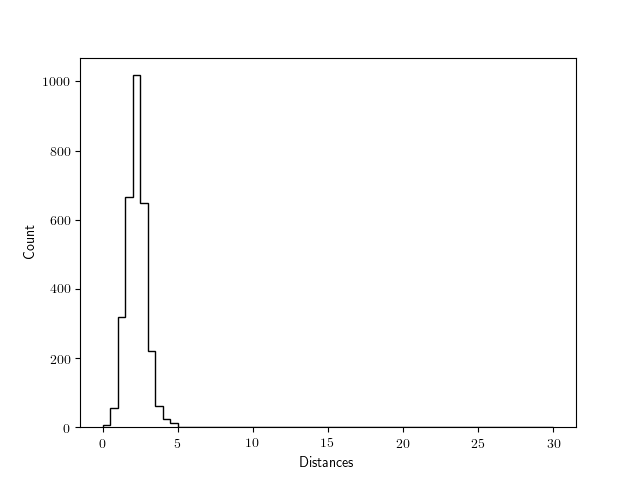}
\caption[Histogram of the distances between neighbouring observables after having resampled with the adaptive method.]{Histogram of the distances between neighbouring observables after having resampled with the adaptive grid-free method (for the distances before sampling, see figure \ref{fig:Distance_histo1}).}
\label{fig:Resampled_Histos2}
\end{figure}

Figure \ref{fig:Resampled_Histos2} shows the result of this algorithm in terms of the distances between observables. The most obvious advantage is that there are no outlier distances at higher bins. The fact that there are no axes in the sampling means that there is no longer any notion of the distances along each axis being different. There is still some spread in distances (between distance $\sim1$ and $\sim4$), which is likely due to border effects, where points (and the resulting observables) may behave differently depending on the dynamics used within the buffer zone.

To summarize: both the sampling based on the average eigenvectors and the adaptive sampling have been shown to exhibit improved homogeneity and isotropy. If we do not require a grid, nor a specific number of points, then the adaptive method is slightly better. The distances cluster slightly tighter, and there are no outlier observables far removed from the rest. Although if a grid system, or fixed number of points, are required, then the eigenvector method is still a vast improvement over the initial sampling.

\section{Implications for Neural Networks}
\label{sec:NN}
Now let us quantify how an optimized sampling improves the learning process of a neural network.

We will be considering a neural network that takes an observable as input (in our case, the power spectrum generated by 21cmFAST, discretized on 12 wavenumber bins and 10 redshift bins), and yields the values of the model parameters as output ($\zeta$, $R_{\mathrm{mfp}}$, $T_{\mathrm{vir}}$). Such a network needs to be trained before it can be used on real observational data. The training set consists of a number of different inputs ($P(k,z)$ in our case), and the associated desired outputs (the corresponding values of $\zeta$, $R_{\mathrm{mfp}}$, and $T_{\mathrm{vir}}$). The hypothesis to be tested is whether training on an optimal sampling of the parameter space will improve the accuracy of the predictions of the resulting network.

For this test we used the Keras framework\footnote{https://keras.io/} to implement a full connected neural network with a single hidden layer. The input-layer contains 120 nodes (12 wavenumber bins $\times$ 10 redshift bins), the hidden layer contains 80 neurons, and the output-layer contains three neurons (one for each parameter).

We trained this network with three different training sets, each containing $\sim500$ points scattered in the same region of the parameter space defined in section \ref{Explored Parameters}. The first training set was that of the na\"ive logarithmic sampling, while the second and third training sets were created using the average-eigenvector and the adaptive grid-free samplings respectively.

\begin{figure*}
\minipage{0.48\textwidth}
  \includegraphics[width=\linewidth]{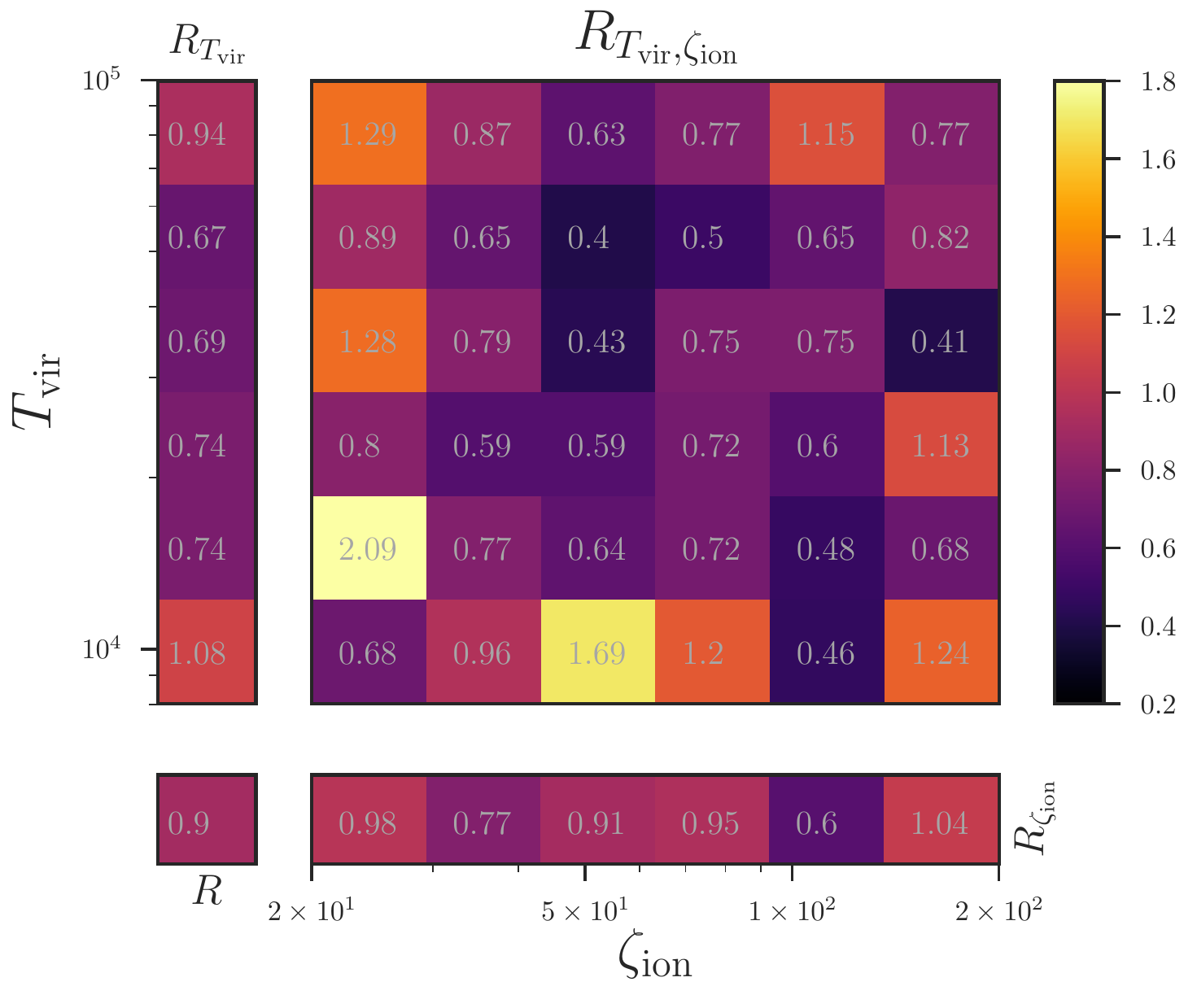}
    \centering
  \textbf{(a)} Average-eigenvector method
\endminipage\hfill
\minipage{0.48\textwidth}%
  \includegraphics[width=\linewidth]{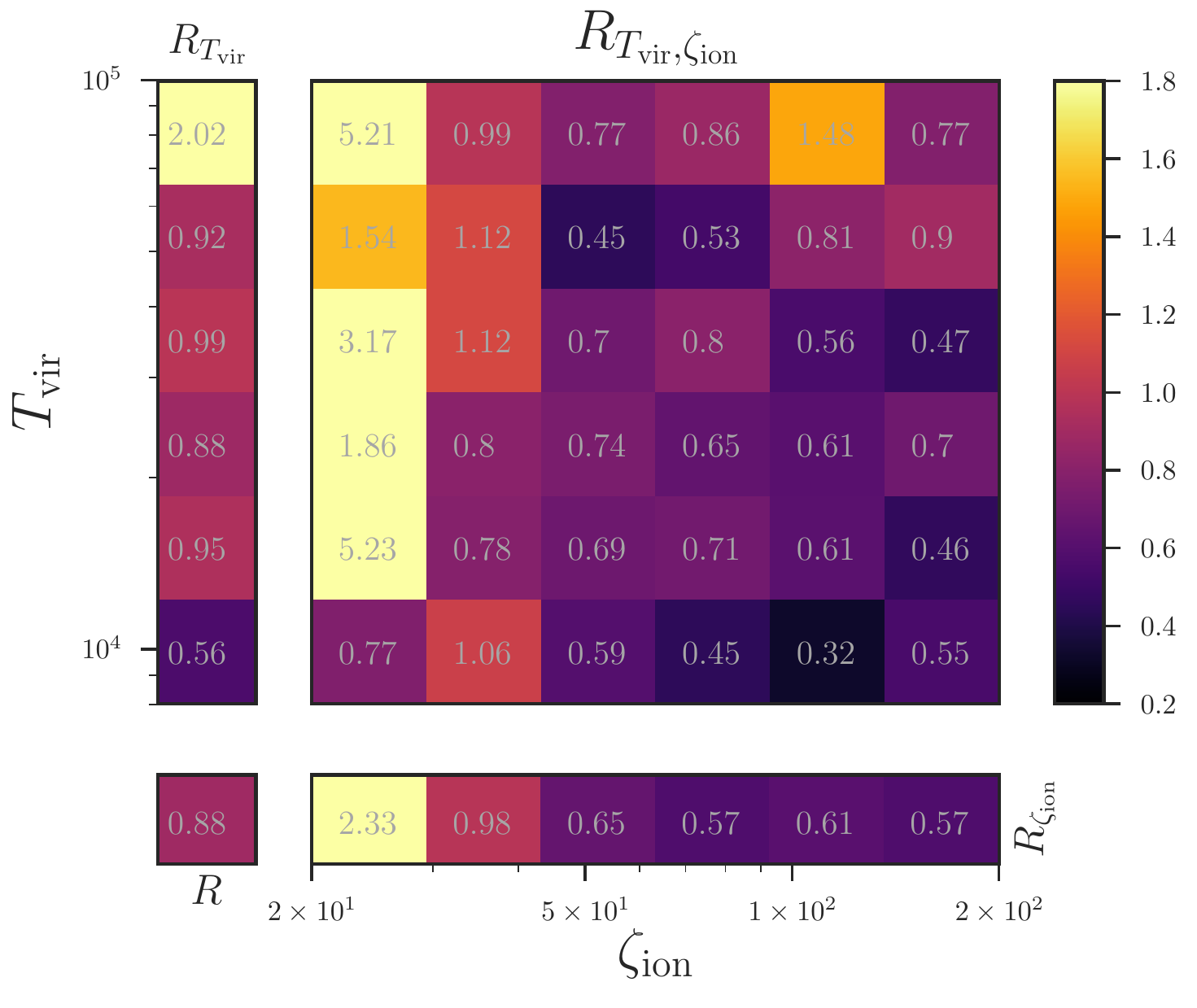}
      \centering
  \textbf{(b)} Adaptive grid-free method
\endminipage
\caption{Parameter reconstruction performance of the parameter space resamplings compared to the initial logarithmic sampling. Each square shows the ratio of the errors on parameter reconstruction for that region ($<1$ indicates the resampling outperforms the logarithmic sampling). The row and column are the ratio of the errors on reconstructing only $\zeta$ and $T_{\textrm{vir}}$, and the square on the bottom left is the overall error ratio across the entire parameter space.} \label{fig:NeuralNetworkPerformance}
\end{figure*}

\subsection{Quantifying Performance}

The accuracy of the network during and after training is evaluated using a different test set consisting of 512 points randomly chosen in the parameter space region. The accuracy of the prediction for sample $j$ in the training set is estimated using a quantity known as the `loss function', and defined in our case as:
\begin{equation}
C_j={1 \over n} \sum_{i=1,n} \left[\log_{10}\left( \theta_{i,j}^{\mathrm{pred}} \over \theta_{i,j}^{\mathrm{true}}   \right)\right]^2
\end{equation}
where $n$ is the number of output parameters (here 3), $\theta_{i,j}^{\mathrm{pred}}$ is the prediction of the network for parameter $\theta_i$ of the sample $j$, and $ \theta_{i,j}^{\mathrm{true}}$ is the true value of parameter $\theta_i$ (the value used by the model to predict the observable for sample $j$). A total cost function $C_{\mathrm{tot}}$ can then be defined for a set of samples by averaging the individual cost functions. Training the network is then reduced to the process of minimizing $C_{\mathrm{tot}}$ by adjusting the neural network weights. This is typically achieved using various forms of gradient descent.  

Using a global cost function defined on the entire test sample was found to discriminate insufficiently between the three choices of training sets resulting from different sampling of the parameter space. 
To remedy this we define $C_{T_{\mathrm{vir}},\zeta}$, a partial cost function defined as the average of the individual cost functions $C_j$ for all samples that fall within some intervals centred on $T_{\mathrm{vir}}$ and $\zeta$. Note that, as the power spectrum is less sensitive to $R_{\mathrm{mfp}}$, we accept all possible values in the partial cost function. Then we define  $R_{T_{\mathrm{vir}},\zeta}$ as the ratio of partial cost functions for an optimized sampling and for the fiducial logarithmic sampling.
Figure \ref{fig:NeuralNetworkPerformance} (a) shows a map of this ratio of the averaged-eingenvector sampling, while Figure \ref{fig:NeuralNetworkPerformance} (b) shows the analagous map for the adaptive sampling. Partial cost functions defined equivalently for a single parameter (thus averaged over all possible values of the other two) are also plotted.
We can see the the overall gain from optimizing the sampling  is small, about $10 \%$ (shown in the bottom left square). However, in both cases, the gain is close to a factor of $2$ if we exclude the boundaries of the parameter space. This is likely due to the fact that the boundary bins in figure \ref{fig:NeuralNetworkPerformance} correspond perfectly to the largest and smallest values of the gridded fiducial logarithmic sampling: the two optimized samplings are at a comparative disadvantage near the boundaries.

A complementary diagnostics is the gain, computed for each parameter separately. This is quantified by computing the $\chi^2$ value for each parameter separately for the test sample. The results are presented in table \ref{tab1} for the whole parameter space region, as well as for a region restriction to an ellipsoid inscribed in the parallelepiped  region.
First, note the improvement in the values compared to \cite{Shimabukuro17}, even with the logarithmic sampling. This is mainly due to the training set being nearly 10 times larger. We also see that for $R_{\mathrm{mfp}}$ the change in gain when using the optimized sampling is negligible when considering the full region, and although it is seen to decrease for the ellipsoid, it changes less than for $\zeta$ and $T_{\textrm{vir}}$. This shows that $R_{\mathrm{mfp}}$ is dominating the error. Comparing the two tables, we verify that restricting the diagnostic to exclude boundary regions, hence the ellipsoid case, increases the gain. Indeed, the logarithmic sampling is favoured by the geometry of the boundaries when considering the whole region. The $\chi^2$ values improve by $15\%$ to $20\%$ within the ellipsoid region.

To summarize, using a training set generated with an informed choice of sampling of the parameter space generally improves the accuracy of the resulting neural network. However, the gain is negatively effected when including a parameter weakly correlated with the inputs ($R_{\mathrm{mfp}}$ in our case), or when predicting parameters near the boundary of the studied parameter space region. As a final caveat, the results presented here have been averaged over $300$ trainings for each type of training samples, as the random initialization of the network weights (a standard practice in neural network training) generates a `noise' in the results comparable to the improvement brought by optimizing the sampling. Thus, optimizing the neural network itself (architecture, error minimization algorithm, regularization, cost function, activation function, etc...) should be carried out first.

\begin{table}
\centering
\begin{tabular}[t]{lccc}
Type of sampling & $\zeta_\mathrm{ion}$ & R$_\mathrm{mfp}$ & T$_{\mathrm{vir}}$ \\
\hline
{\bf Full region:} \\
Log sampling & 0.0293 & 0.0484 & 0.00381 \\
Adaptive & 0.0249 & 0.0518 & 0.00272 \\
Average-eigenvectors & 0.0304 & 0.0464 & 0.00344 \\ 
\hline
{\bf Ellipsoid region:} \\
Log sampling & 0.0252 & 0.0463 & 0.00255 \\
Adaptive & 0.0179 & 0.0406 & 0.00174 \\
Average-eigenvectors & 0.0186 & 0.0390 & 0.00202 \\ 
\end{tabular}
\caption{$\chi ^2$ values computed with the predictions for the test sample for neural networks trained on the three different learning samples. The computation is performed on either the whole parameters space region or a region constricted to a inscribed ellipsoid.}\label{tab1}
\end{table}

\section{Conclusions}
In this work we have outlined a new methodology for sampling a parameter space such that methods for inferring parameter constraints can perform better. The general principle is to generate a sampling of the parameter space that maps onto a homogeneous and isotropic sampling in the space of observables. This requires computing a metric that is similar to the Fisher information metric. Computing the average eigenvectors and eigenvalues associated with this metric, we constructed new step vectors along which to generate a new sampling. This is equivalent to reorienting the grid through rotating and stretching the axes of the parameter space. This technique has been shown to improve the isotropy and homogeneity of the resulting sampling, however some regions of the space of observables still remain in which the observables vary more rapidly than elsewhere.

To remedy this, a second algorithm has been created that, through a relaxation process starting from a random sampling and using the local metric information, produces a grid-free sampling that has been shown to further improve the homogeneity by oversampling the above-mentioned regions of rapidly changing observables.

The two resulting resamplings have been used to train a neural network in parameter reconstruction, and we find that, compared to a neural network trained on a fiducial logarithmic sampling, we are able to reduce the error on parameter reconstruction by 10\% when using the grid reorientation method and 12\% when using the adaptive grid-free method. However, the borders of the chosen parallelepiped region of the parameter space align with the na\"ive sampling geometry. When an inscribed ellipsoid region is considered instead the improvement seen with the optimized sampling is on the order of 30-40\% in terms of the loss function (which is logarithmic) or about 20$\%$ in terms of $\chi^2$. Even if the improvement is moderate, it could be relevant when sampling with full-numerical simulations, for which high computational expense increases the necessity of extracting the most information from the computed samplings.

As mentioned before, the next logical step will be to use the inverse-variance weighted power spectrum such that our metric matches the Fisher information metric. To estimate the usefulness of such a sampling it will be necessary to test it via an efficient neural network trained on noisy signals for which the sources of variance (such as weight initialization) affecting the accuracy of the output are already well under control. Another use will be to use the sampling to derive a discretized prior distribution for the parameters matching Jeffreys prior, and use it for MCMC bayesian inference. Comparing the resulting confidence contours with those derived with a flat prior will give us an idea of how our model choice affect the inference.

Several other developments are possible. A natural question to ask is how this procedure for generating an optimal sampling scales to higher dimensional parameter spaces. One obvious answer is that, to avoid the curse of high dimension computation times, the fiducial sampling used to create the first estimation of the metric will have to be much more sparse. It is then likely that an iterative procedure will be needed, refining the fiducial sampling and the metric estimation in regions where strong variation of the geometry occur. Generally speaking the method we have developed to estimate the metric in the observable space is also a quantitative, useful tool to estimate which observables constrain which parameter efficiently (as well as for which region of the parameter space). Such a comparison could be performed between the power spectrum and the Pixel Distribution Function, for example.

\section*{Acknowledgments}
This work was made in the framework of the French ANR funded project ORAGE (ANR-14-CE33-0016). We also acknowledge the support of the ILP LABEX (under the reference ANR-10-LABX-63) within the Investissements
d'Avenir programme under reference ANR-11-IDEX-0004-02. The simulations were performed on the GENCI
national computing center at CCRT and CINES (DARI grants number 2014046667 and 2015047376).
\bibliographystyle{mnras}
\bibliography{myref}

\appendix

\end{document}